\RequirePackage{ifpdf}
\documentclass[letterpaper]{JHEP3}
\usepackage{amsmath}
\usepackage{epsfig}
\usepackage{subfigure}
\pdfoutput=1

\newcommand{\roughly}[1]{\mathrel{\raise.3ex\hbox{$#1$\kern-0.85em
\lower1ex\hbox{$\sim$}}}}

\def\nn{\nonumber}

\newcommand{\be}{\begin{equation}}
\newcommand{\bee}{\begin{equation}}
\newcommand{\ee}{\end{equation}}
\newcommand{\beea}{\begin{eqnarray}}
\newcommand{\eea}{\end{eqnarray}}
\newcommand{\bea}{\begin{eqnarray}}

\def\nott#1{\setbox0=\hbox{$#1$}                % set a box for #1
   \dimen0=\wd0                                 % and get its size
   \setbox1=\hbox{/} \dimen1=\wd1               % get size of /
   \ifdim\dimen0>\dimen1                        % #1 is bigger
      \rlap{\hbox to \dimen0{\hfil/\hfil}}      % so center / in box
      #1                                        % and print #1
   \else                                        % / is bigger
      \rlap{\hbox to \dimen1{\hfil$#1$\hfil}}   % so center #1
      /                                         % and print /
   \fi}                                         %

\def\uxsl{\hbox{/\kern-.4000em$u$}}
\def\uxslsm{\hbox{\smaller/\kern-.5600em$u$}}
\def\pxpsl{\hbox{/\kern-.5000em$p$}}
\def\epssl{\hbox{/\kern-.5600em$\epsilon$}}
\def\delsl{\hbox{/\kern-.7000em$\nabla$}}
\def\lxpsl{\hbox{/\kern-.5600em$l$}}
\def\kxpsl{\hbox{/\kern-.5600em$k$}}
\def\qxpsl{\hbox{/\kern-.3900em$q$}}

\def\pref#1{(\ref{#1})}
\def\exd{{\rm d}}
\def\ol#1{{\overline{#1}}}

\def\cA{{\cal A}}

\def\cG{{\cal G}}

\def\cL{{\cal L}}

\def\cO{{\cal O}}

\def\cR{{\cal R}}

\def\mfa{{\mathfrak a}}
\def\mfb{{\mathfrak b}}
\def\mfc{{\mathfrak c}}
\def\mfd{{\mathfrak d}}

\def\mfg{{\mathfrak g}}

\def\ssA{{\scriptscriptstyle A}}

\def\ssN{{\scriptscriptstyle N}}

\def\ssS{{\scriptscriptstyle S}}
\def\ssT{{\scriptscriptstyle T}}

\def\PPN{{\scriptscriptstyle PPN}}

\title{Axion Homeopathy: Screening Dilaton Interactions}

\author{C.P.~Burgess${}^{1,2,3}$ and F.~Quevedo${}^{4}$ \\ 

{\it 
${}^1$ Department of Physics \& Astronomy, McMaster University\\ \qquad 1280 Main Street West, Hamilton ON, Canada.\\
${}^2$ Perimeter Institute for Theoretical Physics\\
\qquad 31 Caroline Street North, Waterloo ON, Canada.\\
${}^3$ CERN, Theoretical Physics Department, Gen\`eve 23, Switzerland.\\
${}^4$ DAMTP, University of Cambridge, Wilberforce Road,  Cambridge, CB3 0WA, UK.
}
}

\preprint{CERN-TH-2021-176}

\date{\today}

\abstract{Cosmologically active Brans-Dicke (or dilaton) scalar fields are generically ruled out by solar system tests of gravity unless their couplings to ordinary matter are much suppressed relative to gravitational strength, and this is a major hindrance when building realistic models of light dilatons coupled to matter. We propose a new mechanism for evading such bounds if matter also couples to a light axion, that exploits nonlinear target-space curvature interactions to qualitatively change how the fields respond to a gravitating source. We find that dilaton-matter couplings that would be excluded in the absence of an axion can become acceptable given an additional small axion-matter coupling, and this is possible because the axion-dilaton interactions end up converting the would-be dilaton profile into an axion profile. The trajectories of matter test bodies are then controlled by the much weaker axion-matter couplings and can easily be small enough to escape detection. We call this mechanism Axion Homeopathy because the evasion of the dilaton-coupling bounds persists for extremely small axion couplings provided only that they are nonzero. We explore the mechanism using axio-dilaton equations that are $SL(2,\mathbb{R})$ invariant (as often appear in string compactifications), since for these the general solutions exterior to a spherically symmetric source can be found analytically. We use this solution to compute the relevant PPN parameters, $\gamma_\PPN $ and $\beta_\PPN$, and verify that their deviation from unity can be much smaller than it would have been in the absence of axion-matter couplings and can therefore evade the experimental bounds. }

\begin{document}

\section{Introduction}

Brans-Dicke (BD) theories of gravity have a long, well-motivated history \cite{Jordan, BransDicke, Dicke:1964pna, Brans}. They are perhaps the simplest examples of scalar-tensor theories \cite{ScalarTensorTests} that satisfy the principle of equivalence, and so automatically evade all constraints coming from its very stringent observational tests \cite{EPTests} even if they are light enough to mediate macroscopically long-range forces. They can do so because of their defining feature: the Brans-Dicke scalar couples to matter only as part of a scalar-dependent `Jordan-frame metric', 
\be \label{JFvsEF}
    \tilde g_{\mu\nu} = A^2(\phi) \, g_{\mu\nu} \,,
\ee
that is related to the `Einstein-frame' metric, $g_{\mu\nu}$ (for which the Einstein equations take the standard form) by a Weyl-rescaling factor $A^2(\phi)$. The original Brans-Dicke theory corresponds to the choice
\be
   A(\phi) = e^{\mfg \phi/M_p}
\ee
where $\phi$ is the canonically normalized Einstein-frame scalar field and $\mfg$ is the Brans-Dicke coupling.\footnote{This is related to the traditional Brans-Dicke parameter $\omega$ through the relation $2\mfg^2 =1/(3+2\omega)$.} More general `quasi-Brans-Dicke' models correspond to other choices for $A(\phi)$.  

Although not limited by Equivalence-Principle tests, the strength of BD-matter couplings are constrained to be much weaker than gravity by other, somewhat less sensitive, tests of General Relativity (GR) (see for instance \cite{Will:2014kxa, alphaTests, LOP, Pulsars, Horbatsch:2010hj, GWBounds, DoublePulsar}). The best solar-system tests currently come from the radar-delay measurements performed in the solar system using the Cassini probe \cite{Cassini}, and require that the parameterized post-Newtonian (PPN) metric parameter \cite{PPNDefs} $\gamma_\PPN$, arising in the expansion of $\tilde g_{\mu\nu}$ in powers of $GM/r$, must satisfy
\be \label{CassiniPPN}
    \Bigl| \gamma_\PPN - 1 \Bigr| = \frac{4\mfg^2}{1+2 \mfg^2} < 2.3 \times 10^{-5} \,,
\ee
where $\gamma_\PPN = 1$ is the prediction of GR.

Theorists often revisit this class of models because it is also theoretically well-motivated. Brans-Dicke-like scalars appear naturally in UV complete theories like string theory, where they arise as the dilatons for the various accidental scaling symmetries that are generic in higher-dimensional supergravities \cite{Burgess:2011rv, SUGRAscaleinv, BMvNNQ, GJZ}. These supergravities in turn inherit these scaling symmetries from the perturbative expansions of string theory itself \cite{Burgess:2020qsc}. They also arise in other cosmological models that involve scaling symmetries (accidental or otherwise) \cite{ScaleAnomalyCC}, such as in attempts to describe the current cosmic acceleration in terms of a rolling scalar (quintessence) field \cite{quintessence, Caldwell:1997ii}, and in particular those that rely on exponential scaling potentials \cite{Tracker1, Tracker2, Tracker3, Barreiro:1999zs}. 

Unfortunately most models predict matter-dilaton couplings that are of gravitational strength\footnote{Light axions need not satisfy similar constraints because their pseudoscalar nature can allow them not to couple as strongly to matter.} and so too large to be consistent with \pref{CassiniPPN}. For this reason it is natural to conclude that these scalars should acquire potential energies that give them masses heavy enough to preclude their mediating forces with ranges that can contribute to astrophysics. If so they would be irrelevant for late-universe cosmology. This point of view is also reinforced by the well-known sensitivity of scalar potentials to quantum effects (usually summarized as `technical naturalness' difficulties associated with keeping scalars very light). 

Can Brans-Dicke scalars ever escape these arguments and be relevant to cosmology? So far as masses go, the objection to having a small but technically natural mass can be evaded if the BD scalar is a pseudo-Goldstone boson \cite{pseudoGB} with a shift symmetry.\footnote{Being the Goldstone boson for scale invariance itself is usually insufficient because the strength of scale-breaking effects.} Alternatively, a gravitationally coupled scalar with lagrangian
\be
   \cL = - \sqrt{-g} \left[ \frac{M_p^2}{2} \, \partial_\mu \varphi \, \partial^\mu \varphi + v^4 U(\varphi) \right]
\ee
and generic potential satisfying $U(\varphi_\star) \sim U''(\varphi_\star) \sim \cO(1)$ in the vicinity of a minimum at $\varphi = \varphi_\star$ generically has a mass of order $v^2/M_p$ and it is an old observation \cite{Albrecht:2001xt} that this is automatically extremely light (of order the present-day Hubble scale $H_0$) if there should be a natural explanation why $v^4$ is of order the present-day Dark Energy density ({\it i.e.}~if there were a solution to the cosmological constant problem\footnote{See \cite{CCNo-Scale} for a recent approach to this problem that builds on scale invariance and so indirectly rests on a mechanism like the one we describe to evade solar-system bounds.}).
 
But can gravitational strength dilaton-matter couplings for such light particles escape solar system tests? Over the years several proposals have been made to effectively screen the coupling of a BD field to ordinary matter.  Damour and Polyakov \cite{Damour:1994zq} for instance exploit the fact that matter couplings dominate scalar evolution during cosmology to argue that BD fields should evolve cosmologically to eliminate any matter couplings\footnote{Unfortunately this process is usually too slow for gravitationally coupled scalars, motivating studies with stronger couplings \cite{LOP}.} (see also \cite{Brax:2010gi}). Khoury and Weltman \cite{Khoury:2003aq} proposed the chameleon mechanism in which the presence of matter alters the mass of the BD field, making it large enough within matter to evade the bounds. Other related mechanisms have been also considered (for a review see for instance  \cite{Burrage:2017qrf} and references therein). These screening mechanisms usually assume fairly specific couplings and in particular usually do not apply for scale-invariance dilatons. Despite many attempts, no working mechanism has emerged that applies to the dilaton (or to other moduli fields that arise from UV completions like string theory).

We here introduce a new mechanism that can allow gravitational strength dilaton couplings to evade observational bounds provided the low-energy theory contains an axion partner that combines with the dilaton to have a curved target-space metric. For instance, in string examples the dilaton, $\tau$, and axion, $a$, would be the real and imaginary parts of the complex axio-dilaton field $T = \frac12(\tau + i a)$. Although the dilaton-matter coupling tries to generate an unacceptably large dilaton field, our mechanism uses the dilaton-axion derivative couplings (associated with target-space curvature) to convert the would-be dilaton into a dominantly axion configuration. Provided the axion field has even a very small but non-vanishing coupling to matter (and so must have nonzero derivatives) this mechanism is able to ensure that the resulting field configuration does not much affect test-particle motion (thereby avoiding solar system bounds), and instead mediates weaker spin-dependent interactions which are much less constrained (see for instance \cite{Marsh:2015xka}). The couplings required by our mechanism are generically present in UV completions such as string theory.

We illustrate the physics using the special case of the $SL(2,\mathbb{R})$ invariant axio-dilaton system, for which the field equations external to a spherically symmetric source can be explicitly integrated. We assume when doing so that any potential energy of the axio-dilaton field is negligible compared to the matter density, and we compute the boundary conditions that match the external solutions to the interior solution at the surface of the source ({\it e.g.}~a star) at $r=R$. 

Although our mechanism was motivated by a recent proposal to address the dark energy problem \cite{CCNo-Scale} in which this type of complex axio-dilaton field plays a key role, most of this proposal's bells and whistles are not required for the underlying mechanism described here.

\section{Brans-Dicke scalars}

This section summarizes the bare bones couplings of a Brans-Dicke scalar under the assumption that its scalar potential -- and in particular its mass -- is small enough to be negligible on the scales to be studied. Our interest is in how such scalars impinge on tests of GR in the solar system, and to do so we rederive the PPN parameter that appears in the most dangerous constraints (which the models of \S\ref{sec:AxioDilaton} subsequently evade despite having couplings that at face value should have been ruled out). This section contains largely standard material and so impatient readers and axio-dilaton aficianados can skip this section without loss. 

\subsection{The vanilla model}

Brans-Dicke model is a particular instance of class of scalar-tensor theories that couple a single light scalar $\phi$ to matter and to gravity through a Lagrangian density of the form
\be \label{VanillaL}
    \cL= - \sqrt{- g}\left[ \frac{M_p^2}{2} \, \cR + \frac12 \, (\partial \phi)^2  + V(\phi) \right] + \cL_m(\tilde g_{\mu\nu}, \psi) \,.
\ee
Here $\cR$ is the Ricci scalar built from $g_{\mu\nu}$, while $M_p^{-2} = 8 \pi G $ with $G $ being Newton's constant for universal gravitation and $\psi$ is a representative matter field. The defining feature of this class of models is that $\phi$ couples to matter only through the Jordan-frame metric, defined by
\be \label{JFtoEF}
  \tilde g_{\mu\nu} = A^2(\phi) \; g_{\mu\nu} \,,
\ee
where $g_{\mu\nu}$ is the Einstein-frame metric used in the rest of the lagrangian \pref{VanillaL}. Having matter only couple through $\tilde g_{\mu\nu}$ ensures that the model predicts no preferred-frame effects and so immediately evades all bounds coming from tests of the Equivalence Principle \cite{EPTests}. 

Brans-Dicke theory corresponds to the special case where the scalar potential $V(\phi)$ is negligible and where 
\be
  A(\phi) = \exp\Bigl(\mfg \, \phi/M_p \Bigr) \,,
\ee
for some coupling $\mfg$. For historical reasons $\mfg$ is often traded for the Brans-Dicke parameter $\omega$ defined by\footnote{Notice we use a slightly different normalization for $\phi$ than is common in the gravity literature \cite{Will:2014kxa}.} $2\mfg^2 = 1/(3+2\omega)$. This kind of coupling is theoretically well-motivated and is often encountered in the context of theories with approximate scale invariance for which $\phi$ is called the dilaton and the above assumptions imply that matter fields have a universal $\phi$-dependent mass $m = m_0 \,A(\phi)$.

\subsubsection*{Conserved currents}

With these choices a matter stress-energy can be defined using either metric:
\be \label{TtildeDef}
   \widetilde T^{\mu\nu} := \frac{2}{\sqrt{-\tilde g}} \; \frac{\delta S_m}{\delta \tilde g_{\mu\nu}} 
   \quad \hbox{and} \quad
   T^{\mu\nu} := \frac{2}{\sqrt{- g}} \; \frac{\delta S_m}{\delta g_{\mu\nu}}   \,,
\ee
that are related to one another by
\be \label{TvsTtilde}
    T^{\mu\nu}(x) =  A^6(\phi) \, \widetilde T^{\mu\nu}(x) \,, \quad
    {T_\mu}^\nu = A^4 \, {{\widetilde{T}}_\mu}^\nu \quad \hbox{and} \quad
    T_{\mu\nu} =  A^2 \, \widetilde T_{\mu\nu} \,,
\ee
where ${T_\mu}^\nu = g_{\mu\lambda} T^{\lambda\nu}$ while ${\widetilde T_\mu}^\nu = \tilde g_{\mu\lambda}  \widetilde T^{\lambda\nu}$ and so on. Of these, diffeomorphism invariance ensures that the matter equations of motion alone suffice to imply conservation, $\widetilde D_\mu \widetilde T^{\mu\nu} = 0$, where $\widetilde D_\mu$ is the covariant derivative built using the Christoffel symbol of $\tilde g_{\mu\nu}$. By contrast, diffeomorphism invariance only ensures that the sum of $T^{\mu\nu}$ with the $\phi$-field stress energy is covariantly conserved when the equations of motion for {\it both} $\phi$ and matter are satisfied. 

It is sometimes useful to consider perfect fluids and for these the stress tensors define the pressure and energy density,  
\be \label{PFluid}
    \widetilde T^{\mu\nu} = (\tilde \rho + \tilde p) \, \widetilde U^\mu \widetilde U^\nu + \tilde p \, \tilde g^{\mu\nu} 
    \quad\hbox{and} \quad
      T^{\mu\nu} = (  \rho +  p) \, U^\mu U^\nu + p \,  g^{\mu\nu}  
\ee
where $U^\mu$ and $\widetilde U^\mu$ are the 4-velocities of observers co-moving with the fluid, normalized to satisfy $\tilde g_{\mu\nu} \widetilde U^\mu \widetilde U^\nu = -1$ and $g_{\mu\nu} U^\mu U^\nu = -1$. Chasing through the definitions implies the Einstein-frame and Jordan-frame energy density and pressure are related by
\be \label{prhotrans}
  p = A^4(\phi) \, \tilde p \qquad \hbox{and} \qquad \rho = A^4(\phi) \, \tilde \rho \,,
\ee
and so an equation of state like $w = p/\rho = \tilde p/\tilde \rho$ takes the same form in either frame. 

\subsubsection*{Response to nonrelativistic sources}

Varying the action built from the lagrangian density \pref{VanillaL} leads (in Einstein frame) to the dilaton equation
\be \label{DilEq}
    \Box \phi(x) + \frac{\mfg}{M_p} \,  g_{\mu\nu} T^{\mu\nu}= 0  \,,
\ee
and the trace-reversed Einstein's equation 
\be \label{TTEinst}
  \cR_{\mu\nu} + \frac{1}{M_p^2}\, \partial_\mu \phi \, \partial_\nu \phi + \frac{1}{M_p^2} \left[ T_{\mu\nu} - \frac12 \, g^{\lambda\rho} T_{\lambda\rho} \, g_{\mu\nu} \right] = 0 \,.
\ee

The trace-reversed Einstein equation simplifies in the nonrelativistic limit near flat space, for which the only significant source of stress energy is the energy density $\rho \gg p$ and the only significant source of curvature is $\cR_{tt} \simeq \nabla^2 \Phi$, where $\Phi \in g_{tt}$ is the Newtonian potential. 

In this case, if $\phi$ is time-independent, then the $t$-$t$ component of \pref{TTEinst} simplifies to Poisson's equation
\be \label{PoissonEq}
  \nabla^2 \Phi - \frac{\rho}{2M_p^2} = \nabla^2 \Phi - 4 \pi G  \rho \simeq 0 \,,
\ee
which for a spherically symmetric source of mass $M = \int \exd^3x \; \rho$ implies the usual exterior solution $\Phi = - G  M/r$ where (as usual) an integration constant is chosen by requiring $\Phi$ vanish as $r \to \infty$. 

The dilaton equation \pref{DilEq} in this same limit becomes
\bea \label{DilNR}
   0 = \Box \phi + \frac{\mfg}{M_p} \, g_{\mu\nu} T^{\mu\nu} \simeq \nabla^2 \phi - \frac{\mfg \rho}{M_p} \,.   
\eea   
Defining the dimensionless field $\varphi := \phi/M_p$ we see that $\varphi/(2\mfg)$ satisfies the same equation as does $\Phi$ and so $\varphi \simeq \varphi_\infty + 2\mfg \Phi$. Exterior to the star this implies
\be \label{BDCoulombvsg}
   \varphi = \varphi_\infty - \frac{2\mfg\, G  M}{r} \,.
\ee
What complicates finding similarly explicit solutions to these equations interior to the star is the $\varphi$-dependence that is hidden in $\rho$ due to expressions like \pref{prhotrans} (see for example \cite{Damour:1993hw, Horbatsch:2010hj}).

\subsection{Parameterized post-Newtonian metric}
\label{ssec:BDResponse}

Tests of gravity compare the observed motion of test particles with the motion predicted by the above field configurations. In this theory test particles built from the matter part of the lagrangian move (in the absence of other forces) along the geodesics of the Jordan frame metric, $\tilde g_{\mu\nu}$. They do so because this is the metric that appears in their kinetic term (and so which controls the eikonal approximation for matter fields, whose point-particle limit gives geodesic motion \cite{Will:2014kxa}). 

The parameterized post-Newtonian (PPN) framework \cite{PPNDefs} compares the motion predicted by $\tilde g_{\mu\nu}$ to that obtained in GR order-by-order in powers of $G M/r$ by writing
\bea \label{PPNdefs}
   \tilde g_{\mu\nu} \exd x^\mu \exd x^\nu &=& - \left[ 1 - \frac{2GM}{r} + 2(\beta_\PPN - \gamma_\PPN) \left( \frac{GM}{r}\right)^2 + \cdots \right] \exd t^2 \\
   && \qquad\qquad + \left[ 1 + 2\gamma_\PPN \left(\frac{GM}{r} \right) + \cdots \right] \exd r^2 + r^2 \exd \Omega^2 \,.\nn
\eea
The Schwarzschild solution of General Relativity corresponds to $\gamma_\PPN = \beta_\PPN = 1$ and GR's success in describing solar-system observations using the Cassini probe \cite{Cassini} currently requires the bound \pref{CassiniPPN}.

We next rederive the standard formula for how $\gamma_\PPN$ depends on $\mfg$ so that we can repeat this calculation for the model encountered in \S\ref{sec:AxioDilaton}. Conceptually, there are two reasons why $\tilde g_{\mu\nu}$ differs from the metric one would have had in GR: ($i$) the Weyl factor $A$ in $\tilde g_{\mu\nu} = A^2 g_{\mu\nu}$ causes the Jordan-frame and Einstein-frame metrics to differ, and ($ii$) the Einstein-frame metric $g_{\mu\nu}$ itself solves the Einstein equation with scalar stress energy even exterior to the source (rather than being Ricci flat as in GR). We consider each of these contributions in turn, and show why it is the contribution from the Weyl scaling factor $A(\phi)$ that dominates at leading order in $GM/r$. 

\subsubsection*{Response to $\varphi$ stress-energy}

Consider first the change in the Einstein-frame metric due to the presence of scalar-field stress energy. The field equation \pref{TTEinst} becomes (outside the matter source)
\be \label{TTEinstMAD0}
  \cR_{\mu\nu} +   \partial_\mu \varphi \, \partial_\nu \varphi   = 0 \,,
\ee
and so writing $\varphi = \varphi_\infty + \varphi_1/r$ (with $\varphi_1 = -2 \mfg \, GM $) and adopting the metric 
\be
   \exd s^2 = - e^{2u(r)} \exd t^2 + e^{2v(r)}\exd r^2 + r^2 \exd \Omega^2 \,,
\ee
we have\footnote{Using Weinberg's curvature conventions \cite{GRGrav}.}
\be \label{rrIntEinEQ0}
   \cR_{rr} =  u'' + (u')^2 - u'v'  - \frac{2v'}{r}  = - (\varphi')^2 = - \frac{\varphi_1^2}{r^4} \,,
\ee
while
\be \label{ttIntEinEQ0}
   \cR_{tt} = e^{2(u-v)} \left[- u'' - (u')^2 + u'v'  - \frac{2u'}{r} \right] = 0\,,
\ee
and
\be\label{ththIntEinEQ0}
   \cR_{\theta\theta} = -1 +e^{-2v} \Bigl[ 1 + r(u' - v') \Bigr]  =  0 \,.
\ee
Eliminating $u''$ between \pref{rrIntEinEQ0} and \pref{ttIntEinEQ0} leads to
\be\label{uminusvEQ0}
  -\frac{r}{2} \Bigl[  \cR_{rr} + e^{2(v-u)}\cR_{tt} \Bigr] = u' + v' = \frac{\varphi_1^2}{2r^3} \,,
\ee
which integrates to give
\be\label{uminusvSoln0}
  u + v = -  \frac{\varphi_1^2}{4r^2} \,,
\ee
once the integration constant is fixed by the asymptotic condition $u+v \to 0$ as $r \to \infty$. 

Substituting this into \pref{ththIntEinEQ0} gives a differential equation for $u(r)$ alone that can be approximately solved by seeking a solution as an expansion in powers of $1/r$, leading to 
\be
    u(r) = -\frac{\ell}{r} - \frac{\ell^2}{r^2} + \frac{ 3 \ell \varphi_1^2- 16 \ell^3}{12 r^3} + \cdots \,,
\ee
for integration constant $\ell$. The metric components to leading order in $GM/r$ then are
\be \label{pertmetcompt}
   e^{2u} \simeq  1 - \frac{2\ell}{r} + \frac{\ell \varphi_1^2}{2r^3} + \cO(r^{-4})    \,,
\ee
and
\be \label{pertmetcompr}
   e^{2v} \simeq  1 + \frac{2\ell}{r} + \frac{8\ell^2 - \varphi_1^2}{2r^2} + \frac{16\ell^3 - 3\ell  \varphi_1^2}{2r^3} + \cO(r^{-4})   \,.
\ee
Notice that  \pref{pertmetcompt} and \pref{pertmetcompr} reduce to the usual Schwarzschild solution (expanded in powers of $1/r$) when $\varphi_1 = 0$ with an appropriate choice for $\ell$. Eq.~\pref{BDCoulombvsg} shows that this is to be evaluated at $\varphi_1 = -2\mfg 
\ell$ for Brans-Dicke theory. Because $g_{rr}= e^{2v}$ and $g_{tt} = - e^{2u}$ respectively first deviate from Schwarzschild at order $1/r^2$ and $1/r^3$ the $\varphi_1$-dependence appears with too many powers of $1/r$ to contribute to the key PPN parameters $\gamma_\PPN$ and $\beta_\PPN$ of \pref{PPNdefs}. 

\subsubsection*{Weyl rescaling}

Only the Weyl factor $A$ is therefore relevant for $\gamma_\PPN$ and $\beta_\PPN$. Writing
\be \label{WeylExpansion}
   A = A_\infty \left[1 + \frac{a_1}{r} + \frac{a_2}{r^2} + \cdots \right]
\ee
the Jordan-frame metric becomes
\be
   \tilde g_{\mu\nu} \, \exd x^\mu \, \exd x^\nu = A_\infty^2 \left[1 + \frac{2a_1}{r} + \frac{a_1^2 + 2a_2}{r^2} + \cdots \right]  \Bigl( -e^{2u} \exd t^2  + e^{2v} \exd r^2 + r^2 \exd \Omega^2 \Bigr) \,.
\ee

To preserve the standard form for the metric's angular part we redefine the coordinate
\be
   r \to \hat r := \frac{A(r)}{A_\infty} \; r =  r + a_1 + \frac{a_2}{r} + \cdots 
\ee
so that 
\be
  \frac{1}{r} = \frac{1}{\hat r} + \frac{a_1}{\hat r^2}   + \cdots 
  \quad \hbox{and} \quad
    \exd r =  \exd \hat r \left( 1 +  \frac{a_2 }{ \hat r^2}+ \cdots \right) \,.
\ee
After the additional coordinate rescalings $\tilde t := A_\infty t$ and $\tilde r := A_\infty \hat r$ the metric becomes
\bea
   \tilde g_{\mu\nu} \, \exd x^\mu \, \exd x^\nu 
   &=&  - \left[ 1 - \frac{2(\ell - a_1)A_\infty}{\tilde r} + \cdots \right] \exd\tilde t^2 \nn\\
   && \qquad\qquad +  \left[  1   + \frac{2 (\ell+a_1)A_\infty}{\tilde r} + \cdots \right] \exd \tilde r^2 + \tilde r^2 \exd \Omega^2  \,,\nn
\eea

The coefficient of $1/\tilde r$ in $\tilde g_{tt}$ {\it defines} the gravitational mass of the source if this is only measured by the geodesic motion of bodies in this metric (as is the case for the PPN metric). Defining therefore $GM =(\ell - a_1)A_\infty$ and $\tilde a_1 := a_1 A_\infty$ and $\tilde a_2 := a_2 A_\infty^2$ and so on, we have $- g_{\tilde t \tilde t} = 1 - (2GM/\tilde r)  + \cdots$ while the radial component becomes
\be
   \tilde g_{\tilde r \tilde r}  =  1 + \frac{2(GM + 2\tilde a_1)}{\tilde r} +\cdots \,. 
\ee
Comparing this with \pref{PPNdefs} then implies
\be
  \gamma_\PPN = 1 + \frac{2 \tilde a_1}{GM} =  \frac{\ell + a_1}{\ell - a_1}  \,.
\ee
For instance for a Brans-Dicke scalar we have $A = e^{\mfg \varphi}$ and $\varphi = \varphi_\infty - (2\mfg \ell/r)$ and so $A_\infty = e^{\mfg \varphi_\infty}$ and $a_1 = -2 \mfg^2 \,\ell$, reproducing the familiar result \cite{Will:2014kxa}
\be \label{PPNvsmfa}
  \gamma_\PPN = \frac{1-2\mfg^2}{1+2 \mfg^2} = \frac{\omega + 1}{\omega +2} \,,
\ee
and the last equality trades $\mfg$ for the traditional Brans-Dicke parameter using $(2\mfg)^{-2} = 2\omega + 3$.

Computing $\beta_\PPN$ requires being more careful about $1/r^2$ terms in the scalar field. However if the metric agrees with the Schwarzschild metric to the order found above, then the leading solution \pref{BDCoulombvsg} for $\phi$ also applies relativistically out to order $1/r^2$, but does so in isotropic coordinates -- defined by $\exd s^2 = - e^{2u} \exd t^2 + e^{2w}(\exd \check{r}^2 + \check{r}^2 \exd \Omega^2)$ which for the Schwarzschild geometry implies $e^{2w} = 1 + (2\ell/\check{r})+\cdots$ and $e^{2u} = 1 - (2\ell/\check{r}) + 2(\ell/\check{r})^2+\cdots$. So if $A^2 = A_\infty^2[1 - (\alpha_1/\check r) + (\alpha_2/\check{r}^2) + \cdots]$ when $A = e^{\mfg \phi}$ one finds $\alpha_2 = \frac12 \alpha_1^2$ and $\alpha_1 = -4\mfg^2 \ell$. But when the Einstein-frame metric is Schwarzschild the nonlinear PPN parameter becomes
\be \label{betaPPNform}
 \beta_\PPN = \frac{\ell^2 + \ell \alpha_1 + \frac12 \alpha_2}{(\ell + \frac12 \alpha_1)^2} \,,
\ee
and so gives $\beta_\PPN = 1$ because $\alpha_2 = \frac12 \alpha_1^2$.  

\subsection{Multiple scalars}

The mechanism described here relies fundamentally on the existence of target-space curvature, and so cannot be formulated without having at least two scalars. Consider the action
\be
   S = - \frac{M_p^2}{2} \int \exd^4x \; \sqrt{-g}\; g^{\mu\nu} \Bigl[ \cR_{\mu\nu} + \cG_{ab} (\varphi) \, \partial_\mu \varphi^a \, \partial_\nu \varphi^b \Bigr] + S_m[\tilde g_{\mu\nu}, \psi]
\ee
where again $\tilde g_{\mu\nu} = A^2(\varphi) \, g_{\mu\nu}$. The Einstein equation obtained from this generalizes \pref{TTEinst} to 
\be \label{TTEinstM}
  \cR_{\mu\nu} + \cG_{ab}(\varphi)\, \partial_\mu \varphi^a \, \partial_\nu \varphi^b + \frac{1}{M_p^2} \left[ T_{\mu\nu} - \frac12 \, g^{\lambda\rho} T_{\lambda\rho} \, g_{\mu\nu} \right] = 0 \,,
\ee
where, as before, $T^{\mu\nu}$ is the Einstein-frame stress-energy tensor obtained from $S_m$. 

The scalar equation obtained by varying $\varphi^a$ similarly generalizes \pref{DilEq} to
\be \label{DilEqM}
   \partial_\mu \Bigl[ \sqrt{-g} \; \cG_{ab} \, \partial^\mu \varphi^b \Bigr] + \sqrt{-g} \left[- \frac12 \, \partial_a \cG_{bc} \, \partial_\mu \varphi^b \, \partial^\mu \varphi^c + \frac{\eta_a(\varphi)}{M_p^2} \, g_{\mu\nu} T^{\mu\nu}\right] = 0 \,,
\ee
where the coupling function is defined by
\be
    \eta_a(\varphi) :=  \frac{\partial_a A}{A} \,.
\ee
Eq.~\pref{DilEqM} can be more geometrically written as
\be \label{DilEqMGeo}
      \Box \varphi^a  + \Gamma^a_{bc} \, \partial_\mu \varphi^b \, \partial^\mu \varphi^c + \frac{1}{M_p^2}\, \eta^a(\varphi) \, g_{\mu\nu} T^{\mu\nu} = 0 \,,
\ee
where $\eta^a :=  \cG^{ab} \eta_b$ and 
\be
   \Gamma^a_{bc} := \frac12 \, \cG^{ad} \Bigl[ \partial_b \cG_{cd} + \partial_c \cG_{bd} - \partial_d \cG_{bc} \Bigr] 
\ee
is the Christoffel symbol of the second kind built from the target-space metric $\cG_{ab}$. It is the $\Gamma^a_{bc} \partial \varphi^b \partial \varphi^c$ terms in \pref{DilEqMGeo} that are crucial in the discussions of \S\ref{sec:AxioDilaton}. This term is not usually present in the single-scalar example because it can be eliminated using a field redefinition. 

\subsubsection*{K\"ahler target space}

In order to be more explicit, let us consider a K\"ahler target space typical of the supersymmetric case in which we have complex fields, $z^a$ and $z^{\bar a}$ and the metric is K\"ahler: $G_{ac} = G_{\bar a \bar c} = 0$ while locally $G_{a\bar c} = \partial_a \partial_{\bar c} K$ for some choice of K\"ahler potential $K(z,\bar z)$. For such geometries all components of the Christoffel symbols with mixed $a$ and $\bar a$ indices vanish, leaving as nonzero only the purely holomorphic combination $\Gamma^a_{bc}$ and its complex conjugate:
\be
   \Gamma^a_{bc} = G^{\bar e a} \partial_b G_{c \bar e} = K^{\bar e a} K_{bc\bar e} \,,
\ee
where subscripts as usual denote derivatives. Motivated by \cite{CCNo-Scale} we are also interested in the case where the coupling to matter is controlled by\footnote{Supersymmetry underlies this motivation because the function $K$ controls both the geometry of the scalar target-space metric and the non-minimal coupling of the scalars to the spacetime curvature.}
\be
     A = e^{K/6}  \,,
\ee
in which case the matter-coupling vector has complex components 
\be \label{etaKahler}
  \eta_a = \frac{ \partial_a A }{ A} = \frac{K_a}{6}   \quad \hbox{and} \quad
  \eta^a = K^{\bar c a} \eta_{\bar c} = \frac{K^{\bar c a} K_{\bar c}}{6} \,.
\ee

\subsubsection*{Axio-dilaton example}
\label{sssec:complexfield}

The important yet minimal example used in \S\ref{sec:AxioDilaton} involves two real fields combined into a single complex field $T = \frac12(\tau + i a)$, for which $K = -3  \ln (T + \ol T) = - 3 \ln \tau$. These choices imply 
\be \label{SU11Example}
   G_{\ssT \ol\ssT} = K_{\ssT \ol\ssT} = \frac{3}{(T+\ol T)^2}= \frac{3}{\tau^2} \quad \hbox{and} \quad
   \Gamma^\ssT_{\ssT\ssT} = - \frac{2}{T+\ol T} = -\frac{2}{\tau} \,,
\ee
and the further choice
\be \label{SU11ExampleA}
   A = e^{K/6} = \frac{1}{\sqrt\tau} \quad \hbox{implies} \quad \eta_\ssT = \frac{K_\ssT}{6} = - \frac{1}{2\tau} \quad \hbox{and} \quad  
   \eta^\ssT = - \frac{\tau}{6}\,.
\ee

With these choices --- and choosing non-relativistic matter (for which $g_{\mu\nu} T^{\mu\nu} = - \rho$) --- eq.~\pref{DilEqMGeo} then becomes
\be \label{DilEqMGeoTNR}
      \Box T  - \frac{2}{\tau} \,\partial_\mu T \, \partial^\mu T + \frac{\tau}{6 M_p^2} \, \rho = 0 \,,
\ee
whose real and imaginary parts then give
\be \label{DilEqMGeoTNRi}
      \Box a  - \frac{2}{\tau} \,\partial_\mu \tau \, \partial^\mu a = 0 
\ee
and
\be \label{DilEqMGeoTNRr}
      \Box \tau  - \frac{1}{\tau} \Bigl[ \partial_\mu \tau \, \partial^\mu \tau - \partial_\mu a \, \partial^\mu a \Bigr] + \frac{\tau}{3 M_p^2} \, \rho = 0 \,.
\ee

Notice that constant $a$ solves these equations and switching in this case to the new variable $\tau = e^{\zeta\varphi}$ [for which $\tau^{-1} \Box \tau = \zeta\,\Box \varphi + \zeta^2 \partial_\mu \varphi \, \partial^\mu \varphi$] implies \pref{DilEqMGeoTNRr} reduces to the Brans-Dicke form $\Box \varphi = -\rho/(3\zeta M_p^2)$ (as in \pref{DilNR}). As elaborated below, canonical normalization corresponds to $\zeta = \sqrt{2/3}$, and so comparing to \pref{DilNR} shows that the Brans-Dicke coupling is $\mfg = - 1/\sqrt{6}$. 

\section{Axio-Dilaton}
\label{sec:AxioDilaton}

The previous section introduces a simple axio-dilaton model involving a complex field $T = \frac12(\tau + i a)$ for which the dilaton-matter coupling obtained for $\tau$ when $a$ is constant is $\mfg = - 1/\sqrt{6} \simeq - 0.41$. At face value, using $\gamma_\PPN = (1-2\mfg^2)/(1+2\mfg^2)$ then implies $|\gamma_\PPN -1| = \frac12$, which is much too large to satisfy the observational constraint \pref{CassiniPPN}. 

In this section we show how even extremely small matter couplings to the axion $a$ can allow this size of a matter-dilaton coupling to escape solar system bounds. This happens because the nonlinear $\tau$-$a$ derivative couplings modify the prediction \pref{PPNvsmfa} for $\gamma_\PPN$, and do so by having the matter-$\tau$ coupling largely produce an external $a$ profile rather than a $\tau$ profile. This evades solar-system bounds because the axion field influences test-body motion much less efficiently than does the dilaton.

We continue to assume throughout this section that the axion and dilaton masses are negligible for solar-system applications (but return briefly in \S\ref{sec:Conclusions} to comment on possible implications of an axion mass).

\subsection{Symmetries and field equations}

Consider therefore the special case of the axio-dilaton described in \S\ref{sssec:complexfield} in somewhat more detail. The model's lagrangian density is
\be \label{axiodilL}
   \cL  =  - \sqrt{-g} \; M_p^2 \left[ \frac{\cR}2 + \frac{3\, \partial^\mu \ol T \, \partial_\mu T}{(T+\ol T)^2}\right] + \cL_m  
    = - \sqrt{-g} \; M_p^2 \left[ \frac{\cR}2 + \frac34  \left( \frac{\partial^\mu \tau \, \partial_\mu \tau + \partial^\mu a \, \partial_\mu a}{\tau^2} \right) \right] + \cL_m \,,  
\ee
where $T = \frac12(\tau + i a)$ and we assume the matter couples to $\tau$ only through the Jordan-frame metric $\tilde g_{\mu\nu} = A^2(\tau) \, g_{\mu\nu}$ with $A = \tau^{-1/2}$. We do allow the possibility that matter can also couple directly to the axion independent of $\tilde g_{\mu\nu}$, though only very weakly. Writing $\tau = e^{\zeta \varphi}$ and demanding the kinetic term in \pref{axiodilL} become $-\frac12\, M_p^2 \sqrt{-g}\; (\partial \varphi)^2$ is what determines $\zeta = \sqrt{2/3}$, as quoted earlier.

Specialized to non-relativistic sources these choices give the following axio-dilaton field equations
\be \label{DilEOM}
      \Box \tau  - \frac{1}{\tau} \Bigl( \partial_\mu \tau \, \partial^\mu \tau - \partial_\mu a \, \partial^\mu a \Bigr) + \frac{\tau \, \rho }{3 M_p^2}= 0 \,,
\ee
and
\be \label{AxEOM}
      \Box a  - \frac{2}{\tau} \,\partial_\mu \tau \, \partial^\mu a + \frac{\tau^2  \cA}{3 M_p^2}  = 0 \,,
\ee
where $\rho = - g_{\mu\nu}T^{\mu\nu}$ is used for non-relativistic matter with
\be
   T^{\mu\nu} := \frac{2}{\sqrt{-g}} \; \frac{\delta S_m}{\delta g_{\mu\nu}} \quad \hbox{and} \quad \cA := \frac{2}{\sqrt{-g}} \; \frac{\delta S_m}{\delta a} \,.
\ee
The trace-reversed Einstein equations are given by \pref{TTEinstM}, specialized to the above choices:
\be \label{TTEinstMAX}
  \cR_{\mu\nu} + \frac{3}{4\tau^2} \Bigl(  \partial_\mu \tau \, \partial_\nu \tau +  \partial_\mu a \, \partial_\nu a \Bigr) + \frac{1}{M_p^2} \left[ T_{\mu\nu} - \frac12 \, g^{\lambda\rho} T_{\lambda\rho} \, g_{\mu\nu} \right] = 0 \,.
\ee

\subsubsection{$SL(2,\mathbb{R})$ invariance and conservation laws}

These equations prove to be relatively simple to integrate, largely because of the number of conservation laws that they admit. In particular notice that the non-matter part of the action is invariant under the following $SL(2,\mathbb{R})$ transformation
\be \label{SL2Rdef}
   T = \frac12(\tau + i a) \to \frac{\mfa T -i \mfb}{i \mfc T + \mfd} 
\ee
where the four real parameters $\mfa$ through $\mfd$ satisfy the constraint $\mfa \mfd - \mfb \mfc = 1$. Notice that these three real parameters include as special cases the axionic shift symmetry $a \to a -2 \mfb$ (when $\mfa = \mfd = 1$ and $\mfc = 0$) and scale invariance $T \to \mfa^2 T$ (when $\mfd = 1/\mfa$ and $\mfb = \mfc = 0$).

Noether's theorem implies there must be three conserved currents, and these can be taken to be 
\be
  J_\ssA^\mu = \frac{\partial^\mu a}{\tau^2} \qquad \hbox{(axion shift symmetry)}
\ee
\be
   J_\ssS^\mu = \frac{\partial^\mu \tau}{\tau} + \frac{a \,\partial^\mu a}{\tau^2} \qquad \hbox{(scaling symmetry)}
\ee
as well as
\be
   J_\ssN^\mu = \frac{(\tau^2 - a^2)}{\tau^2} \; \partial^\mu a - \frac{2a}{\tau} \; \partial^\mu \tau
   \qquad \hbox{(nonlinear symmetry)}  \,,
\ee
where the last conservation law corresponds to the nonlinear transformation $\delta T = - i \epsilon T^2$ obtained when $\mfa = \mfd = 1$ while $\mfb = 0$ and $\mfc = \epsilon \ll 1$. As is straightforward to verify, direct differentiation together with use of the field equations \pref{DilEOM} and \pref{AxEOM} implies these currents are all conserved in the absence of sources
\be
   D_\mu J^\mu_\ssA 
   = - \frac{\cA}{3M_p^2} \,, \quad
     D_\mu J_\ssS^\mu 
     = - \frac{(\rho +a \,\cA)}{3M_p^2} \quad \hbox{and} \quad
     D_\mu J_\ssN^\mu 
     = \frac{(a^2 - \tau^2) \cA + 2a \,\rho}{3M_p^2} \,.
\ee

\subsubsection*{Spherically symmetric source}

We next record the field equations and conservation laws for spherically symmetric solutions, for which $\tau = \tau(r)$ and $a = a(r)$. Denoting differentiation with respect to $r$ by primes, the field equations \pref{DilEOM} and \pref{AxEOM} reduce to
\be \label{DilEOMrad}
      \tau'' + \frac{2\tau'}{r}  - \frac{(\tau')^2}{\tau}  + \frac{(a')^2}{\tau}  + \frac{\tau\, \rho }{3 M_p^2} = 0 \,,
\ee
and
\be \label{AxEOMrad}
      a'' + \frac{2\,a'}{r}  - \frac{2 \, a' \tau'}{\tau}  + \frac{\tau^2 \cA}{3 M_p^2}  = 0 \,,
\ee
while the three conservation laws become
\be \label{Aconsrad}
   \left[ r^2 \left( \frac{a'}{\tau^2} \right) \right]' = - \frac{r^2 \cA}{3M_p^2} \qquad \hbox{($J_\ssA^\mu$ conservation)}\,,
\ee
\be \label{Sconsrad}
   \left[ r^2 \left( \frac{\tau'}{\tau} + \frac{a \, a'}{\tau^2}\right) \right]' = - \frac{r^2}{3M_p^2} (\rho +a \,\cA) \quad \hbox{($J_\ssS^\mu$ conservation)}
\ee
and
\be \label{Nconsrad}
   \left\{ r^2 \left[ \frac{(\tau^2 - a^2)a'}{\tau^2}  - \frac{2a\, \tau'}{\tau} \right] \right\}' =   \frac{r^2}{3M_p^2} \Bigl[ (a^2 - \tau^2) \cA + 2\, a \,\rho \Bigr]
   \quad \hbox{($J_\ssN^\mu$ conservation)} \,.
\ee

The conservation laws \pref{Aconsrad} and \pref{Sconsrad} provide useful information when they are integrated within the interior of the source, using the boundary condition $a'(0) = \tau'(0) = 0$ at the center of the matter distribution at $r = 0$ that is required by spherical symmetry.  Assuming $a$ and $\tau$ themselves remain bounded at $r = 0$ we find expressions for the radial field derivatives just exterior to the edge of the source (which we imagine occurs at $r = R$):
\be \label{Aconsradint}
    \left( \frac{a'}{\tau^2} \right)_{r=R} = - \frac{1}{3M_p^2R^2}\int_0^R \exd r \; r^2 \cA(r) \,,
\ee
and
\be \label{Sconsradint}
    \left( \frac{\tau'}{\tau} + \frac{a \, a'}{\tau^2}\right)_{r=R} = - \frac{1}{3M_p^2R^2} \int_0^R \exd r\; r^2 \Bigl[\rho(r) +a(r) \,\cA(r) \Bigr]\,.
\ee
The implication of the third conservation law is explored in the next section.

\subsection{Spherically symmetric exterior solutions}

We next explicitly integrate the axio-dilaton field equations to obtain their general solution external to the gravitating source.

\subsubsection{Trivial axion solution}

For the purposes of making contact to the Brans-Dicke special case it is worth first exploring the limit $\cA = 0$. In this case the axion equation \pref{AxEOMrad} is trivially solved by a constant axion field $a = a_0$ and the dilaton equation \pref{DilEOMrad} then becomes
\be \label{DilEOMradtrivcan}
      \varphi'' + \frac{2\varphi'}{r}  + \frac{\rho}{3\zeta M_p^2}  = 0 \,,
\ee
where $\tau(r) := e^{\zeta\varphi(r)}$. 

This last form is the usual Brans-Dicke result, and is solved in the exterior region (where $\rho = 0$) by the usual Coulomb-like solution found in previous sections
\be \label{BDCoulomb}
  \varphi = \varphi_\infty + \frac{\varphi_1}{r} \,,
\ee
with integration constants $\varphi_\infty$ and $\varphi_1$. In this case the above conservation laws degenerate into the single independent condition
\be \label{DilatonChg0}
   \left[ r^2 \left( \frac{\tau'}{\tau}\right) \right]' =\zeta \Bigl(r^2 \varphi'\Bigr)' = - \frac{r^2\rho}{3 M_p^2} 
\ee
that implies the boundary condition
\be \label{BDbc}
    \varphi'(R) = - \frac{1}{3\zeta M_p^2R^2} \int_0^R \exd r \; r^2 \rho = - \frac{2GM }{3\zeta R^2}  
\ee
	just exterior to the source, where $M = 4\pi \int_0^R \exd r\; r^2 \rho(r)$ is the mass in the absence of gravitational back-reaction. Comparing to \pref{BDCoulomb} determines the integration constant $\zeta\varphi_1 = \frac23 GM$ and comparing this to \pref{BDCoulombvsg} implies $\mfg = - 1/(3\zeta) = - 1/\sqrt6$, as before. We wish to similarly determine how the boundary conditions \pref{Aconsradint} and \pref{Sconsradint} fix the integration constants in the more general case where $\cA \neq 0$.

\subsubsection{Full axio-dilaton solutions}

The first step to this end is to find the general exterior solutions to identify its integration constants and the existence of so many conservation laws allows this to be done more explicitly than is usually possible. In the absence of sources $\cA = \rho = 0$ and eqs.~\pref{DilEOMrad} and \pref{AxEOMrad} become
\be  
      \tau'' + \frac{2\tau'}{r}  - \frac{(\tau')^2}{\tau}  + \frac{(a')^2}{\tau}  = 0 \quad \hbox{and} \quad
      a'' + \frac{2\,a'}{r}  - \frac{2 \, a' \tau'}{\tau}   = 0 \,.
\ee

The conservation laws provide immediate first integrals of these equations. In particular, when $\rho = \cA = 0$ \pref{Aconsrad} and \pref{Sconsrad} trivially integrate to give
\be \label{Aconsradvac}
    \frac{a'}{\tau^2} = \frac{C_\ssA}{r^2} 
\ee
and
\be \label{Sconsradvac}
 \frac{\tau'}{\tau} = \frac{C_\ssS}{r^2} - \frac{a \, a'}{\tau^2} = \frac{C_\ssS - a \, C_\ssA}{r^2} \,.
\ee
Using these to eliminate $\tau'$ and $a'$ in the third conservation law \pref{Nconsrad} then gives an algebraic condition relating $\tau$ and $a$ 
\be \label{Nconsradvac}
  \frac{C_\ssN}{r^2}  =  \frac{(\tau^2 - a^2)a'}{\tau^2}  - \frac{2a\, \tau'}{\tau}    =  \frac{(\tau^2 + a^2) C_\ssA}{r^2}  - \frac{2a C_\ssS }{r^2} \,.
\ee

\begin{figure}[t]
\begin{center}
\includegraphics[width=140mm,height=70mm]{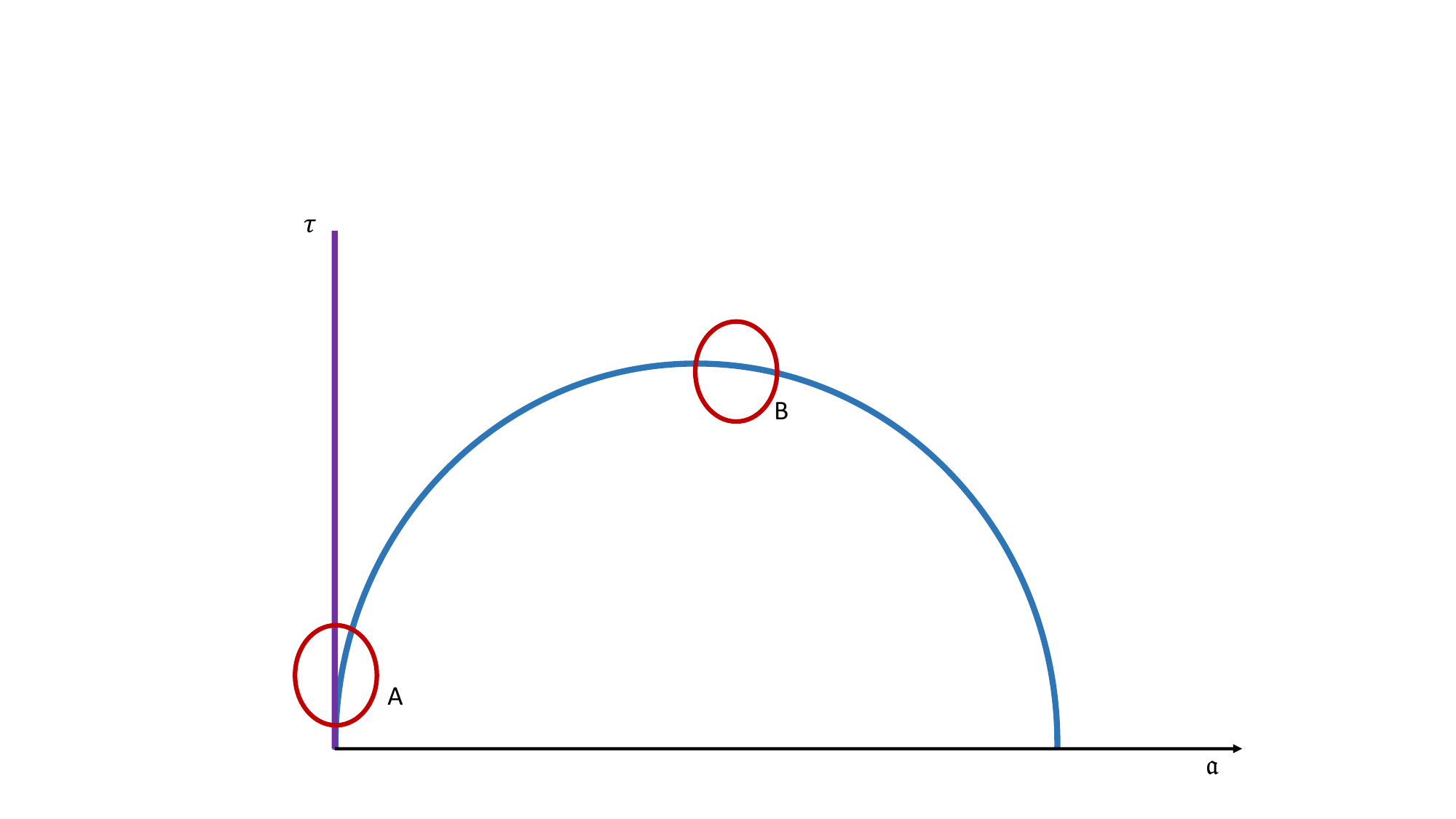} 
\caption{Plot of semicircular axio-dilaton trajectories in the $\tau$--$a$ plane, together with an example of the vertical straight line trajectory obtained in the special case $a' = 0$. Region A shows how the semicircle closely approximates the vertical straight line in the limit of very large radius. Region B shows the screening regime in which the dilaton is suppressed for any small but finite radius.} \label{Fig:Semicircle} 
\end{center}
\end{figure}

For later purposes it is noteworthy that the character of the solutions to these equations when $C_\ssA = 0$ can differ qualitatively from those that arise when $C_\ssA \neq 0$. To see why, notice that if $a' = C_\ssA = 0$ then equations \pref{Sconsradvac} and \pref{Nconsradvac} imply
\be
  \frac{\tau'}{\tau} = \frac{C_\ssS}{r^2} \quad \hbox{and} \quad
  \frac{2a \tau'}{\tau} = -\frac{C_\ssN}{r^2} \,,
\ee
and so $C_\ssN + 2a C_\ssS = 0$. This is the trivial solution given in \pref{BDCoulomb} for which $\tau$ varies monotonically with $a$ fixed, corresponding to a vertical line when drawn in the $a$-$\tau$ plane. But when $C_\ssA \neq 0$ eq.~\pref{Nconsradvac} instead implies
\be \label{Semicircle}
   \tau^2 + \left( a - \alpha\right)^2 = \beta^2 \,,
\ee
where
\be
    \alpha := \frac{C_\ssS}{C_\ssA} \quad \hbox{and} \quad 
    \beta^2 := \left( \frac{C_\ssS}{C_\ssA}\right)^2 + \frac{C_\ssN}{C_\ssA} \,.
\ee
This says that solutions sweep out circles of radius $\beta$ centered at the point $a = \alpha$ on the $a$-axis of the $a$-$\tau$ plane (or -- more properly -- semi-circles in the upper-half $a$-$\tau$ plane since $\tau > 0$: see Fig.~\ref{Fig:Semicircle}). 

Now comes the crucial point: retrieving Brans Dicke theories in the $C_\ssA \to 0$ limit corresponds mathematically to the observation that any vertical line is well approximated as the arc of a circle of infinitely large radius (as in region $A$ of Fig.~\ref{Fig:Semicircle}). But that does not mean that all arcs of that same circle look like vertical lines (such as region $B$ of Fig.~\ref{Fig:Semicircle}). For any nonzero $C_\ssA$ {\it no matter how small} there are always places on the circle where the dilaton $\tau$ reaches a maximum and then decreases and these parts of the solutions do not resemble the vertical straight lines of Brans Dicke theory at all. 

\begin{figure}[t]
\begin{center}
\includegraphics[width=140mm,height=70mm]{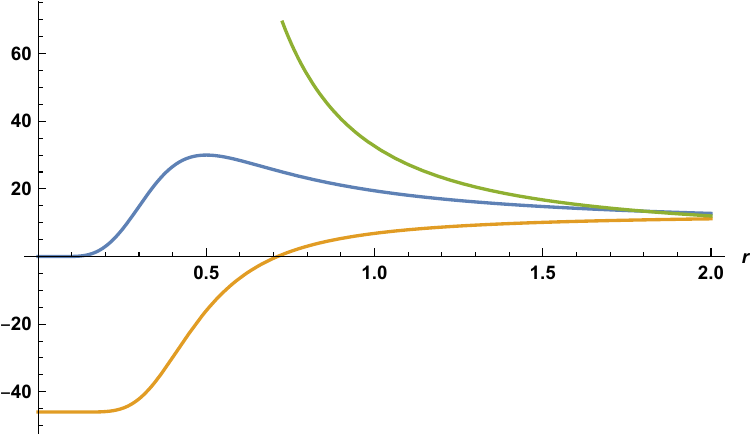} 
\caption{Sample dilaton (blue) and axion(orange) profiles as functions of $r$. Also shown is the dilaton profile (green) in the absence of axions ({\it i.e.}~when $a'=0$).} \label{Fig:DilatonAxion} 
\end{center}
\end{figure}

The explicit $r$-dependence of these solutions is found by using \pref{Semicircle} to eliminate $\tau$ from \pref{Aconsradvac}, leading to
\be \label{Aconsradvacaonly}
   a' = \frac{C_\ssA \tau^2}{r^2} = \frac{C_\ssA}{r^2} \left[  \left( \frac{C_\ssS}{C_\ssA}\right)^2 + \frac{C_\ssN}{C_\ssA} -  \left( a - \frac{C_\ssS}{C_\ssA}\right)^2 \right]
   = \frac{C_\ssA}{r^2} \Bigl[ \beta^2 -  \left( a - \alpha \right)^2 \Bigr] \,.
\ee
This integrates to give
\be
   \int \frac{\exd a}{\beta^2 -(a - \alpha)^2} = C_\ssA \int \frac{\exd r}{r^2}  \,,
\ee
and the result is only consistent with $\tau^2 > 0$ when $\beta^2 > 0$, in which case
\be \label{avsrexact}
    a(r) = 
    \alpha - \beta \tanh X \quad \hbox{with} \quad X(r) := \frac{\beta\gamma}{r} + \delta\,,
\ee
and $\gamma := C_\ssA$. Using this in \pref{Semicircle} then gives
\be \label{tauvsrexact}
   \tau(r) = \frac{\beta}{\cosh X(r)} \,.
\ee
We obtain in this way the general spherically symmetric solutions to the vacuum field equations, parameterized by the four integration constants $\alpha$, $\beta$, $\gamma$ and $\delta$. Fig.~\ref{Fig:DilatonAxion} shows representative plots showing how $\tau$ and $\alpha$ evolve with $r$ and how this evolution differs from what would have occuring for $\tau$ in the absence of the target-space axion-dilaton coupling.

Two of the integration constants are determined by the conservation laws \pref{Aconsradint} and \pref{Sconsradint} which become
\be \label{Aconsradint2}
   \gamma = C_\ssA = R^2 \left( \frac{a'}{\tau^2} \right)_{r=R} = - \frac{1}{3M_p^2}\int_0^R \exd r \; r^2 \cA(r) \,,
\ee
while
\be \label{Sconsradint2}
   \gamma \alpha = C_\ssS = R^2 \left( \frac{\tau'}{\tau} + \frac{a \, a'}{\tau^2}\right)_{r=R} = - \frac{1}{3M_p^2} \int_0^R \exd r\; r^2 \Bigl[\rho(r) +a(r) \,\cA(r) \Bigr]\,.
\ee
These combine to give a formula for the radial derivative of $\tau$ at the source's surface:
\be \label{logtauprimeBC}
  \left( \frac{\tau'}{\tau} \right)_{r=R} = - \frac{\gamma[a(R) - \alpha]}{R^2} = - \frac{1}{3M_p^2R^2} \int_0^R \exd r\; r^2 \Bigl\{\rho(r) + [a(r) - a(R)] \,\cA(r) \Bigr\} \,.
\ee
The third conservation law \pref{Nconsrad} gives $C_\ssN$ in terms of the integration constants, with
\be \label{Nconsradchk}
   C_\ssN= r^2 \left[ \frac{(\tau^2 - a^2)a'}{\tau^2}  - \frac{2a\, \tau'}{\tau} \right]  =  \gamma (\beta^2 - \alpha^2) 
   \quad \hbox{($J_\ssN^\mu$ conservation)} \,,
\ee
but does not provide independent information beyond what is already contained in \pref{Semicircle}.

A simple estimate for the size implied for these integration constants can be found by assuming the axion source density has the same profile -- but different normalization -- as the energy density: $\cA(r) = \varepsilon \, \rho(r)$, where our interest is in the limit $|\varepsilon | \ll 1$. In this case the boundary conditions \pref{Aconsradint2} through \pref{logtauprimeBC} become  
\be \label{Aconsradint22}
   \gamma = C_\ssA = - \frac{2\varepsilon GM}{3} \,,
\ee
while
\be \label{Sconsradint22}
   \gamma \alpha = C_\ssS = - \frac{1}{3M_p^2} \int_0^R \exd r\; r^2 \rho(r)\Bigl[1 + \varepsilon \, a(r)  \Bigr] \simeq - \frac{2GM}{3}\,,
\ee
and so
\be \label{logtauprimeBC2}
  \left( \frac{\tau'}{\tau} \right)_{r=R} =  - \frac{1}{3M_p^2R^2} \int_0^R \exd r\; r^2 \rho(r) \Bigl\{1+ \varepsilon \,[a(r) - a(R)] \Bigr\} \simeq - \frac{2GM}{3R^2}\,.
\ee

The boundary condition \pref{logtauprimeBC} shows that the dilaton `charge' takes its usual form \pref{DilatonChg0} if $a(R)$ agrees with the weighted average (weighted by $\cA$) of $a$ inside the source. This never occurs if $\cA$ has a definite sign, because then $\gamma$ must have the opposite sign as must $a'(r)$ for all $r$ interior to the source. But this also implies that $a$ is monotonic within the source and so $a(R)$ cannot agree with the source average of $a(r)$.  

Similarly, $\gamma \alpha$ and $\tau'/\tau$ (at $r=R$) must both be negative if $\cA$ is negligible relative to $\rho$ in eqs.~\pref{Sconsradint2} and \pref{logtauprimeBC}. The value of $\tau'/\tau$ at $r=R$ is also negative whenever $\cA$ has a definite sign (regardless of whether or not it dominates $\rho$), because this dictates the sign of $a'$ and so also fixes the sign of $a(r) - a(R)$ in such a way that ensures $[a(r) - a(R)]\cA$ is positive. And once we know that $\tau'(R) < 0$ the fact that $\tau$ evolves along a circle centered on the $a$-axis implies that $\tau(r)$ is monotonically decreasing outside the source as well. 

Eqs.~\pref{Aconsradint2} and \pref{Sconsradint2} show that the boundary conditions at $r = R$ dictate the two integration constants $\alpha$ and $\gamma$. The remaining two integration constants are instead determined by the values of the fields at spatial infinity: $a(r \to \infty) = a_\infty$ and $\tau(r \to \infty) = \tau_\infty$ which the solutions \pref{avsrexact} and \pref{tauvsrexact} imply are given by 
\be \label{Asymptoticatau}
   a_\infty = \alpha - \beta \tanh \delta \quad\hbox{and}\quad
   \tau_\infty = \frac{\beta}{\cosh \delta} = \sqrt{\beta^2 - (a_\infty - \alpha)^2} \,.
\ee
In general the integration constants are therefore determined as follows: $\beta$ and $\delta$ are fixed using \pref{Asymptoticatau}, and so are usually specified by the physics of the environment (such as by matching to the homogeneous fields bequeathed by earlier epochs of cosmology). This boundary condition identifies the circle on which these asymptotic fields must lie. The precise position on this circle as a function of $r$ is then set once $\alpha$ and $\gamma$ are fixed by the properties of the local gravitating source through eqs.~\pref{Aconsradint2} and \pref{Sconsradint2}.

For future use we also record the solution's asymptotic approach to the limiting forms \pref{Asymptoticatau}, applicable when $|\beta\gamma/r| = \frac23\,\beta \varepsilon GM/r \ll 1$ (where we use \pref{Aconsradint22} to evaluate $\gamma$):
\bea \label{LargerAsympt}
   \tau &=& \frac{\beta}{\cosh X} = \frac{\beta}{\cosh \delta}\left[ 1 - \frac{\beta \gamma}{r} \, \tanh \delta + \cdots \right]   \nn\\
   a - \alpha &=&- \beta \tanh X = -\beta  \left[ \tanh \delta + \frac{\beta \gamma/r}{\cosh^2 \delta} + \cdots \right] \,.
\eea
Notice in particular that the dilaton `charge' -- defined as the coefficient of $1/r$ in the expression for $\ln\tau$ in the far-field regime -- is proportional to the parameter combination $\beta \gamma \tanh \delta =  \frac23\,\beta \varepsilon \tanh \delta \, GM$. This shows that the combination $\frac13\,\beta \varepsilon \tanh \delta$ is what would be interpreted as $\mfg$ in the absence of the axion field, and as we see below is what controls the size of the PPN parameter $\gamma_\PPN$. As the next section shows, observable effects in tests of gravity are suppressed as either $\gamma$ ({\it i.e.}~$\varepsilon$) or $\delta$ tend to zero.  

\subsection{Gravitational response}

We now collect all the threads: having such an explicit exterior solution allows us to compute the metric response for this system and thereby determine its implications for the motion of matter test-particles. 

In principle the presence of the direct matter-axion coupling embodied by $\cA(r)$ means that motion of matter particles differs from the treatment described in \S\ref{ssec:BDResponse}, because direct axion-induced forces can prevent them from moving along geodesics of the Jordan-frame metric $\tilde g_{\mu\nu}$. We choose as our particular focus here the limit where $|\cA| \ll \rho$ so that these direct axion-related forces are negligible. This is a fairly natural limit given that the pseudoscalar nature of the axion makes its individual couplings to atoms not sum as coherently in macroscopic matter as does the energy density. 

What is interesting is that despite this a tiny but nonzero $\cA$ nevertheless significantly changes how matter responds to the dilaton, because the dilaton-axion couplings can divert the external field of a source away from the dilaton and towards the axion. 

To see how this works we therefore follow \S\ref{ssec:BDResponse} and assume that $\cA$ is small enough that direct axion forces are negligible in which case test particles move along the geodesics of the Jordan frame metric, $\tilde g_{\mu\nu}$. Tests of gravity then test whether these differ from the geodesics predicted by General Relativity. Just as in \S\ref{ssec:BDResponse} there are two reasons why $\tilde g_{\mu\nu}$ differs from GR: ($i$) the Weyl factor $A$ in $\tilde g_{\mu\nu} = A^2 g_{\mu\nu}$ causes the Jordan- and Einstein-frame metrics to differ, and ($ii$) the Einstein-frame metric $g_{\mu\nu}$ solves the Einstein equation with axiodilaton stress energy rather than being Ricci flat as in GR. 

Consider first the second of these effects. The Einstein-frame metric's response to scalar stress-energy is expressed by the field equation \pref{TTEinstM} with $\rho = \cA = 0$:
\be \label{TTEinstMAD}
  \cR_{\mu\nu} + \frac{3}{4\tau^2} \Bigl[ \partial_\mu \tau \, \partial_\nu \tau + \partial_\mu a \, \partial_\nu a \Bigr]  = 0 \,.
\ee
Specialized to $a = a(r)$ and $\tau = \tau(r)$ only the component $\cR_{rr}$ is nonzero, satisfying
\be \label{TTEinstrrMAD}
   \cR_{rr} = - \frac{3}{4} \left[ \frac{(\tau')^2 + (a')^2}{\tau^2} \right] 
   = - \frac34 \left[ \frac{C_\ssS^2 - 2 a C_\ssS C_\ssA + C_\ssA^2(a^2 + \tau^2)}{r^4} \right] 
   = - \frac{3\gamma^2 \beta^2}{4r^4} \,.
\ee
This uses the conservation laws \pref{Aconsradvac} and \pref{Sconsradvac} together with the semi-circle condition \pref{Semicircle} and the definitions $\gamma = C_\ssA$ and $\beta^2 = (C_\ssS/C_\ssA)^2 + (C_\ssN/C_\ssA)$. Notice how the scalar stress energy is suppressed by the axion coupling $\gamma = -\frac23 \varepsilon GM$ -- see \pref{Aconsradint22} -- and so can be made arbitrarily small by dialing down the matter-axion coupling $\varepsilon$.

What is important is that eq.~\pref{TTEinstrrMAD} is an exact expression for the axio-dilaton solution everywhere exterior to the star, and does not invoke a large-$r$ expansion. This means that from the point of view of the metric the stress energy for the full axio-dilaton solution is precisely the same as for a Brans-Dicke field -- {\it c.f.}~eq.~\pref{rrIntEinEQ0} -- with the replacement $\varphi_1^2 \to 3\beta^2 \gamma^2$. Repeating the arguments of \S\ref{ssec:BDResponse} then show that within a parameterized post-Newtonian expansion the leading corrections to the Einstein-frame metric far from the source first arise at order $1/r^2$ for $g_{rr}$ and at order $1/r^3$ for $g_{tt}$ and so are too small to contribute to the PPN parameters $\gamma_\PPN$ and $\beta_\PPN$. 

This leaves only the contribution $(i)$ listed above; the contribution of the Weyl factor in the expression $\tilde g_{\mu\nu} = A^2 g_{\mu\nu}$. In the present case the large-$r$ expansion of eq.~\pref{LargerAsympt} shows that the far-field position-dependence of the Weyl factor predicted by the axio-dilaton solutions is 
\be \label{Avstauandr}
   A^2 = e^{K/(3M_p^2)} = \frac{1}{\tau} = \frac{\cosh X}{\beta} 
   = \frac{\cosh \delta}{\beta} \left[ 1+ \frac{\beta\gamma}{r} \tanh \delta + 
   \frac{\beta^2\gamma^2}{2r^2}  + \cdots \right] \,.
\ee
Writing $A^2 = A^2_\infty \left[1 - ({\alpha_1}/{r}) +(\alpha_2/r)^2+ \cdots \right]$ allows the results of the Brans-Dicke analysis in \S\ref{ssec:BDResponse} to be carried over in whole cloth subject only to the replacement
\be \label{a1vsdelta}
   \alpha_1 = - \beta\gamma\, \tanh \delta  =  \frac{2\varepsilon \,\beta \ell}{3}\, \tanh\delta \quad\hbox{and} \quad
   \alpha_2 = \frac{\beta^2 \gamma^2}{2}  = \frac{2\varepsilon^2\beta^2  \ell^2}{9}  \,,
\ee
where use of the flat-space solutions assumes the metric is written in isotropic coordinates, as described in \S\ref{ssec:BDResponse}. As in earlier sections we write the Einstein-frame metric as $-g_{tt} = 1 - (2\ell/r) + \cdots$ while reserving $2GM$ for the coefficient of $1/r$ in the Jordan-frame metric.

In particular the PPN parameters controlling the leading corrections to the metric components $\tilde g_{t t}$ and $\tilde g_{\tilde r \tilde r}$ become
\be
  \gamma_\PPN  = \frac{\ell - \frac12 \, \alpha_1}{\ell + \frac12 \, \alpha_1} = \frac{3 - \varepsilon \beta \tanh\delta}{3 + \varepsilon \beta \tanh\delta} 
\ee
and
\be
  \beta_\PPN  = \frac{\ell^2 + \ell \alpha_1 + \frac12 \, \alpha_2}{(\ell + \frac12 \, \alpha_1)^2} = 1 + \frac{ \varepsilon^2 \beta^2}{9(\cosh \delta + \frac13\,\varepsilon \beta \sinh\delta)^2} \,.
\ee
What is clear is that both $\gamma_\PPN -1$ and $\beta_\PPN-1$ can be made as small as desired by making $\cA$ (and so also $\varepsilon$) sufficiently small, since $\gamma$ (and so also $\varepsilon$) is completely determined by $\cA$ through the boundary condition \pref{Aconsradint2}. This suppression can also be enhanced if $\delta$ -- whose value is controlled by the boundary condition \pref{Asymptoticatau} at spatial infinity -- is also small, such as occurs if $a_\infty$ does not to differ much from the value of $\alpha$ implied by \pref{Sconsradint2}.

Once the PPN parameters are suppressed in this way all solar system tests are evaded at once. This is because it is these same PPN parameters that also enter into lunar laser-ranging (LLR) bounds (through their contribution to the Nordvedt effect) and other solar system tests \cite{Will:2014kxa}.\footnote{One might also worry about gravitational binding-energy effects introducing small $\phi$-dependence into the Jordan-frame mass for sources of order $\delta m(\phi) \sim \cO(\mfg^2 GM/R)$, because $G$ depends on $\phi$ in Jordan frame (or equivalently because masses depend on $\phi$ in Einstein frame). However such small terms only introduce deviations  from geodesics of the JF metric in test-body motion by amounts proportional to even smaller local gradients of $\phi$ within the solar system, making them too small to be measureable at present.}

\section{Conclusions}
\label{sec:Conclusions}

In summary, we propose a concrete screening mechanism for Brans-Dicke scalar fields that can allow gravitational-strength couplings to be consistent with solar system tests of GR. In order to be effective the mechanism requires the existence of additional scalar fields with small but nonzero matter couplings and nontrivial target-space derivative couplings to the Brans-Dicke field.

We explore the mechanism in an explicit model consisting of an axio-dilaton whose target space enjoys an $SL(2,\mathbb{R})$ symmetry. The conservation laws associated with this symmetry allow explicit solutions to the field equations outside a spherically symmetric source, and this allows us to determine the exact axion and dilaton profiles. We find these profiles to have several noteworthy features.

\begin{itemize}
\item Test-particle motion dominantly responds to the Jordan-frame metric, and outside the source this differs in two ways from the metric in standard General Relativity. It first differs because of the Weyl factor $A$ that relates the Einstein and Jordan frame metrics in eq.~\pref{JFvsEF}. The second difference arises because the Einstein-frame metric responds to the stress-energy in the axio-dilaton field exterior to the source. Of these, the former dominates because the curvature due to the presence of axio-dilaton stress-energy appears in the metric at order $1/r^2$ and so is too small to contribute to the post-Newtonian parameter $\gamma_{PPN}$.

\item The contribution to $\gamma_\PPN$ coming from the Weyl factor comes through the coefficient $a_1$ given in \pref{a1vsdelta}. This coefficient can be determined in terms of the constants of integration and -- provided there is a small but nonzero coupling of axion to matter -- it can easily be small enough to satisfy the experimental constraint \pref{CassiniPPN}.

\item The mechanism works because the target space curvature converts the field outside the star from dilaton to axion despite the dilaton-matter coupling being much larger than the axion-matter coupling. The axion field is not detected in solar-system tests because its coupling to matter is assumed to be small enough that direct axion-induced forces are negligible and because it does not appear in the Weyl factor, to which test-particle motion is mostly sensitive.

\item The Brans-Dicke limit of vanishing axion-matter coupling is a subtle one and is illustrated in Fig.~\ref{Fig:Semicircle}. In general the solutions trace out semicircles in the $\tau$--$a$ plane, centered on a point along the $a$-axis. This degenerates to a vertical line in the limit of vanishing axion-matter coupling; a solution that becomes indistinguishable from the semicircle in the limit of extremely large circle radius (see region $A$ of Fig.~\ref{Fig:Semicircle}). But for any finite radius, no matter how large, the other parts of the semicircle (such as region $B$ in Fig.~\ref{Fig:Semicircle}) do {\it not} resemble a vertical straight line. This is why small nonzero axion couplings can change the field response relative to zero coupling in such a homeopathic way.
\end{itemize}

We believe our mechanism can help open up the exploration of cosmologies driven by axio-dilaton fields by providing a mechanism for them to evade solar-system constraints despite their gravitational-strength dilaton couplings. Although this does not in itself also solve the technical naturalness problems (small scalar masses and small vacuum energies) associated with such cosmologies, it provides a necessary ingredient once these problems are addressed. Indeed our mechanism emerged from parallel work aimed at these naturalness problems in which the axio-dilaton structure used here plays a central role (details of which can be found in companion papers \cite{CCNo-Scale, LowESugra}). 

There is clearly much more that can be done to develop and better understand the homeopathic screening mechanism. GR is tested in a great many ways these days, and it is worthwhile establishing which of these tests is most sensitive to the presence of an axio-dilaton with the assumed matter couplings. 

Although not explored here, the scenario of \cite{CCNo-Scale} also allows (but does not require) the axion to acquire a mass by being eaten by a dark photon through the Higgs mechanism. In particular, the same parameters that describe the Dark Energy density easily can give a mass of order $10^{-12}$ eV, corresponding to a Compton wavelength of order $10^5$ km whose presence could also have distinctive implications for observations in the solar system. 

The cosmology of such a model also becomes of great interest, and \cite{CCNo-Scale} provides first steps towards identifying successful cosmology. Intriguingly, because the resulting model causes particle masses to vary as the dilaton evolves, differences between its value at recombination and now provide an opportunity to help ameliorate the Hubble tension \cite{HubbleTension} via the mechanism described in \cite{Sekiguchi:2020teg} (see also \cite{Zahn:2002rr, Cyr-Racine:2021alc}).

\section*{Acknowledgements}
We thank Clare Burrage, Ed Copeland, Shanta de Alwis, Danielle Dineen, Nemanja Kaloper, Luis Lehner, Francesco Muia, Sergey Sibiryakov and Adam Solomon for many helpful conversations. CB's research was partially supported by funds from the Natural Sciences and Engineering Research Council (NSERC) of Canada. Research at the Perimeter Institute is supported in part by the Government of Canada through NSERC and by the Province of Ontario through MRI.  The work of FQ has been partially supported by STFC consolidated grants ST/P000681/1, ST/T000694/1.

\end{document}